\documentstyle[preprint,aps,eqsecnum]{revtex}
\begin{document}
\draft
\preprint{Alberta Thy-25-94 hep-th/9412010}
\title{Finite particle creation in 1+1 dim. compact in
space}
\author{D.\ J. Lamb\cite{email}, A.\ Z. Capri }
\address{Theoretical Physics Institute, \\
Department of Physics,
University of Alberta,\\
Edmonton, Alberta T6G 2J1, Canada}
\date{\today}
\maketitle
\begin{abstract}
In this paper we calculate the massive particle creation as seen by a
stationary
observer in a $1+1$ dimensional spacetime compact in space. The
 Bogolubov transformation relating the annihilation
and creation operators between two spacelike surfaces is calculated.
The particle creation, as observed by a stationary observer who moves from the
first spacelike surface to the second
is then calculated, and shown to be finite, as is expected for a spacetime with
finite spatial volume.
\end{abstract}

\section{Introduction}
In the last couple of years there has been renewed interest in the problems
associated with defining particles on a curved manifold \cite{recent}. Much of
this renewed
interest shares common ground in the interpretation of the ``particles"
detected by Rindler observers. The standard analysis as outlined in Birrell and
Davies \cite{1.1} involves relating the annihilation and creation operators of
two different quantizations of a field. One quantization is based on the entire
spacetime while the other is based on coordinates which only cover the Rindler
wedge. As has been shown by Silaev and Krustalev \cite{1.2}, the  boundary
conditions which one is now forced to impose on the boundaries of the wedge are
responsible for the frequency mixing between these modes which is then
interpreted as particle creation. Their analysis compares these standard
calculations for Rindler observers to those done if one quantizes a field in
one half of $1+1$ Minkowski space and compares this to the quantization done
over
the entire spacetime. Indeed many calculations have been done calculating the
Bogolubov transformation relating the operators from two different
coordinatisations, one which covers the entire spacetime and one which only
covers a portion.

 Recently some particle creation calculations have been done by determining
how an observer's particle definition evolves with time \cite{1.3}. The purpose
of this paper is to show that the total particle creation for an expanding
$1+1$ dimensional spacetime compact in space, is finite. This is in agreement
with Fulling's analysis \cite{fulling} as we are dealing with a finite volume
of
space. An earlier calculation for an infinite volume of space yielded
inconclusive
results for the total particle creation which was presumably infinite
\cite{1.4}.
This particle creation as interpreted through a nonvanishing $\beta(n,s)$ in
the Bogolubov
transformation,  drops off faster than any inverse power of the momenta $n,s$
 which implies that the particle creation if finite and that the Bogolubov
transformation is unitarily implementable.

This calculation follow's the procedure of Capri and Roy \cite{Caproy92} which
is
very similar to the procedure advocated by Massacand and Schmid \cite{eth}.
Both these approaches are based on a coordinate independent approach where the
geometry determines the foliation one should use to quantize the field. This
preferred direction of time is given along a normal to the spacelike surface
consisting of those spacelike geodesics which are orthogonal to the observers
4-velocity. In this way the construction only depends on the geometry, the
observers position, and the tangent to the observer's worldline.
\section{The Model}
The model we investigate is that of a compact $1+1$ dimensional
spacetime described by the metric,
\begin{equation}
ds^2=dT^2-e^{\lambda  T}R^2d\theta^2. \ \ \ \ \ \ \ \ \ \
0 \leq \theta < 2\pi
\label{1.1}
\end{equation}
To follow the prescription mentioned above we first
must calculate the geodesics. The first integrals of the
geodesic equations are:
\begin{equation}
\frac{d\theta}{ds}=\frac{c_1}{Re^{\lambda  T }}\ \ \ \ \
\frac{dT}{ds}=\sqrt{\epsilon +
\frac{c_1^2}{e^{\lambda  T}}}
\label{1.2}
\end{equation}
where $\epsilon=\pm 1$ depending on whether the geodesic is timelike or
spacelike respectively.

The preferred coordinates on the hypersurface of instantaneity are constructed
using a 2-bein of orthogonal basis vectors based at $P_0$, the observer's
position. We choose these vectors to be,
\begin{equation}
e_0(P_0)=(1,0)\ \ \ e_1(P_0)=(0,\frac{1}{Re^{\frac{\lambda  T_0}{2}}}),
\label{1.3}
\end{equation}
in this way $e_0(P_0)$ is tangent to the observer's worldline at $P_0$.

To construct a spacelike geodesic which is orthogonal to the observer's
world line it is required that,
\begin{equation}
\frac{dT}{ds}\left|_{P_0}=0\right. \ \ \ {\rm which\ \ \ implies} \ \ \
\frac{c_1^2}{e^{\lambda  T_0}}=1
\label{1.4}
\end{equation}
The preferred coordinates on the spacelike hypersurface are chosen to be
Riemann coordinates based on the observer's position $P_0=(T_0,\theta_0)$.
The point $P_1=(T_1,\theta_1)$ is the point at which a timelike geodesic
``dropped" from an arbitrary point $P=(T,\theta)$ intersects the spacelike
hypersurface orthogonally. The Riemann coordinates $\eta^\alpha$ of the point
$P_1$ are given by,
\begin{equation}
sp^\mu=\eta^\alpha e_\alpha^\mu (P_0)
\label{1.5}
\end{equation}
where $s$ is the distance along the geodesic $P_0-P_1$ and $p^\mu$ is the
tangent vector, at $P_0$, to the geodesic connecting $P_0$ to $P_1$.
These equations can be solved for the $\eta^\alpha$ using the orthogonality
of $p^\mu$ to $e_0(P_0)$ and the identity $e_\alpha^\mu e_{\beta \mu}=
\eta_{\alpha \beta}$ (Minkowski metric) to give,
\begin{equation}
\eta^0=s p^\mu e_\mu^0(P_0) \ \ \ \ \eta^1=-s p^\mu e_\mu^1(P_0)
\label{1.6}
\end{equation}
The surface of instantaneity $S_0$ is just the surface $\eta^0=0$ and
the preferred spatial coordinate $x^1=\eta^1$ is,
\begin{equation}
x^1=s \frac{c_1}{\sqrt{e^{\lambda  T_0}}}
\label{1.7}
\end{equation}
where $s$ is the geodesic distance between the points $P_0$ and $P_1$.
The direction of time is given by the normal to this spacelike hypersurface.
The preferred time coordinate $t$ for the point $P$ is given by the proper
distance along the timelike geodesic connecting $P$ to $P_1$. This timelike
geodesic is also determined by (\ref{1.2}) with $\epsilon = -1$ and a different
 choice of integration constant, $b_1$. The condition that the geodesic
 connecting $P$ to $P_1$ is normal to the spacelike hypersurfaceat $P_1$
requires that
\begin{equation}
\sqrt{1+\frac{b_1^2}{e^{\lambda  T_1}}}\sqrt{\frac{c_1^2}
{e^{\lambda  T_1}}-\frac{c_1^2}{e^{\lambda  T_0}}}=\frac{b_1c_1}{e^{\lambda
T_1}}
\label{1.8}
\end{equation}

The metric can now be calculated in terms of the preferred coordinates
$(t,x^1)$ by calculating $(T(t,x^1),\theta(t,x^1))$. To calculate these
dependances we use the above equation for $x^1$ (\ref{1.7}) and also
calculate the change in the coordinate $\theta$ along the spacelike
and timelike geodesics which connect $P_0$ to $P$,
\begin{equation}
R\theta = R\theta_0 + \int_{T_0}^{T_1}dT\frac{c_1}{e^{\lambda  T_1}}
 \left(\frac{c_1^2}{e^{\lambda  T}}- \frac{c_1^2}{e^{\lambda
T_0}}\right)^{-\frac{1}{2}} +  \int_{T_1}^{T}dT'\frac{b_1}{e^{\lambda
T'}}\left(1+\frac{b_1^2}{e^{\lambda  T'}}
 \right)^{-\frac{1}{2}}
\label{1.9}
\end{equation}
and
\begin{equation}
t=\int_{T_1}^{T}dT'\left(1+\frac{b_1^2}{e^{\lambda  T'}} \right)^{-\frac{1}{2}}
\label{1.10}
\end{equation}
By performing the above integral for $\theta^1$ and inverting the $t$ integral
 one is left with the coordinate transformations
\begin{eqnarray}
e^{\frac{\lambda }{2}(T-T_0)}&=&\sinh(\frac{\lambda  t}{2})+
\cosh(\frac{\lambda  t}{2})\cos(\frac{\lambda  x^1}{2}) \nonumber \\
\frac{R\lambda }{2}(\theta -\theta_0)e^{\lambda \frac{T}{2}} &=&
-\cosh(\frac{\lambda  t}{2})\sin({\frac{\lambda  x^1}{2}})
\label{1.11}
\end{eqnarray}
In terms of the preferred coordinates $(t,x^i)$ the metric now has the form,
\begin{equation}
ds^2=dt^2-\cosh^2(\frac{\lambda  t}{2})(dx^1)^2.
\label{1.12}
\end{equation}
The range of $x^1$ is $0 \leq x^1 < \frac{4\pi}{\lambda }$.
To write this in a more convenient form we introduce the angular
coordinate $\alpha=\frac{\lambda  x^1}{2}$ which covers the range
$0 \leq \alpha< 2\pi$. In terms of this angular coordinate the metric
takes the form
\begin{equation}
ds^2=dt^2-\frac{4}{\lambda ^2}\cosh^2(\frac{\lambda  t}{2})d\alpha^2
\label{1.13}
\end{equation}

\section{Modes and Initial conditions}
In the coordinates constructed above the minimally coupled massive
Klein Gordon equation is,
\begin{equation}
\partial_t^2\phi+\frac{1}{\sqrt{g}}\partial_t(\sqrt{g})\partial_t\phi
+\frac{1}{\sqrt{g}}\partial_1(\sqrt{g}g^{11})\partial_1\phi+m^2\phi=0
\label{2.1}
\end{equation}
To quantize a scalar field on the $t=0$ surface we now define the positive
frequency modes. The positive frequency modes are defined as those which
satisfy the initial conditions,
\begin{equation}
\phi_k^{+}(t,{\bf x})\left|_{t=0}\right.=A_k(0,\alpha)\ \ \ {\rm and }\ \ \
\partial_t(\phi_k^{+}(t,\alpha))\left|_{t=0}\right.=-i\omega_k(0) A_k(0,\alpha)
\label{2.2}
\end{equation}
Where $A_k(t,\alpha)$ are the instantaneous eigenmodes of the spatial part of
the Laplace-Beltrami operator, and $\omega_k(t)^2$ are the corresponding
eigenvalues.
\begin{equation}
\left[\frac{1}{\sqrt{g}}\partial_1\left(\sqrt{g}g^{11}\partial_1 \right) +m^2
\right]A_k(t,\alpha)=\omega_k^2(t)A_k(t,\alpha)
\label{2.3}
\end{equation}
Henceforth we just write $\omega_k$ for $\omega_k(0)$.
Due to the simple form of $g_{\mu\nu}$ at $t=0$ the eigenmodes and eigenvalues
take on the simple form,
\begin{eqnarray}
A_k(0,\alpha)&=&e^{i{\frac{2k}{\lambda }\alpha}} \\ \nonumber
\omega_k^2(0)&=&k^2 + m^2.
\label{2.4}
\end{eqnarray}
Near the surface $t=0$ the second term of (\ref{2.1}) vanishes to $O(t^2)$,
this implies that the initial conditions for the time dependence of the field
are also good to $O(t^2)$.
We impose periodic boundary conditions on $A_k(0,\alpha)$ to
choose a self adjoint extension for the differential operator on the
left hand side of
(\ref{2.3}). This requires that $\frac{2k}{\lambda }=s$
where $s$ is an integer.

To impose the initial conditions we need a complete set of modes for the
entire wave operator. Because the wave equation is separable we look for
solutions of the form $f_s(t)e^{is\alpha}$. The differential equation satisfied
by the $f_s(t)$ is then,
\begin{equation}
\partial_t^2 f_s(t) + \frac{\lambda }{2} \tanh(\frac{\lambda  t}{2})\partial_t
f_s(t) + \left(\frac{s^2\lambda ^2}{4} {\rm sech}^2(\frac{\lambda  t}{2}) + m^2
\right)f_s(t)=0
\label{2.5}
\end{equation}
The positive frequency modes are those whose time part satisfies the above
differential equation and the initial conditions,
\begin{equation}
f_s(0)=1 \ \ \ {\rm and }\ \ \ \ \dot{f}_s (0) = -i\omega_k.
\label{3.5}
\end{equation}
The positive frequency modes are given in terms of hypergeometric
functions $H(a,b,c,g(t))$ by
\begin{equation}
\phi_s^+(t,\alpha)=e^{is\alpha}{\rm cosh}(\frac{\lambda
t}{2})^{s}\left\{\right.
H(\alpha,\beta,\frac{1}{2},-\sinh^2(\frac{\lambda  t}{2}))
-\left. i\frac{2\omega_s}{\lambda }
\sinh(\frac{\lambda
t}{2})H(\alpha+\frac{1}{2},\beta+\frac{1}{2},\frac{3}{2},-\sinh^2(\frac{\lambda
t}{2}))\right\}
\label{2.6}
\end{equation}
where
\begin{eqnarray}
\alpha &=& \frac{s}{2}+\frac{1}{4}+\frac{i}{4\lambda }\sqrt{16m^2-\lambda ^2}
\nonumber \\
\beta  &=& \frac{s}{2}+\frac{1}{4}-\frac{i}{4\lambda }\sqrt{16m^2-\lambda ^2}
\nonumber \\
\omega_s  &=& \sqrt{(\frac{\lambda  s}{2})^2+m^2}. \nonumber \\
\label{2.7}
\end{eqnarray}

We can now write out the field which has been quantized on surface $1$ as,
\begin{equation}
\Psi_1=\sum_{s=-\infty}^{s=\infty} \frac{1}{\sqrt{2\omega_s}}\left\{
\phi_s^+(t,\alpha)a_1(s)+\phi_s^{+ \ast}(t,\alpha)a_1^\dagger(s)\right\}
\label{3.8}
\end{equation}

\section{Particle Creation}
To investigate particle creation in the model universe as observed by an
observer stationary with respect to the original coordinates $(T,\theta)$
we calculate the Bogolubov transformation relating the annihilation and
creation operators from two different surfaces of quantization that the
observer passes through. To calculate the coefficients of this transformation
we equate the same field from two different quantizations on a common surface,
\begin{equation}
\Psi_1(t,\alpha)=\Psi_2(t'(t,\alpha),\alpha'(t,\alpha)).
\label{4.1}
\end{equation}
Here $\Psi_1(t,\alpha)$ is the field written out explicitly in (\ref{3.8}) and
$\Psi_2(t',\alpha')$ is the same field which has been quantized on a second
surface $t'=0$. The ``second" field is therefore quantized for the same
observer as the first but at some later time $T'_0$ with $\theta_0=\theta_0'$.
All the physics of the
observations made by this observer are determined by the functions
$t'(t,\alpha)$, $x'(t,\alpha)$ and the derivatives of these functions with
respect to $t$. In this way the geometry of the spacetime via the coordinate
independent prescription we have used, determines the spectrum of created
particles.

We calculate the Bogolubov transformation by ``matching" the
field and its first derivative with respect to $t$ at $t=0$.
\begin{eqnarray}
a_1(s)&=&\frac{i}{(2\pi)}\frac{1}{\sqrt{2\omega_s}}\int_0^{2\pi}d\alpha
e^{-is\alpha}\left\{i\omega_s\Psi_1(0,x)
-\left(\partial_t\Psi_1(t,\alpha)\right)\left|_{t=0}\right. \right\}
\nonumber \\
   &=&\frac{i}{(2\pi)}\frac{1}{\sqrt{2\omega_s}}\int d\alpha
e^{-is\alpha}\left\{i\omega_s\Psi_2(t'(0,\alpha),\alpha'(0,\alpha))
-\left(\partial_t \Psi_1(t'(t,\alpha), \alpha'(t,\alpha))
\right)\left|_{t=0}\right. \right\} \nonumber \\
\label{4.2}
\end{eqnarray}
Using this equation, we can write out the Bogolubov transformation in the form
\begin{equation}
a_2(n)=\sum_s \alpha(n,s)a_1(s) + \beta(n,s)a_1^{\dagger}(s).
\label{4.3}
\end{equation}
The spectrum of created particles is determined by $\left|\beta(n,s)
\right|^2$.

Writing out $\beta(n,s)$ explicitly we find it has some interesting properties
due to it's dependence on the inverse relations $t'(t,x)$,$x'(t,x)$,
\begin{equation}
\beta(n,s)=\frac{-i}{2\pi}\int\! d\alpha
\frac{e^{-in\alpha}}{\sqrt{4\omega_n\omega_s}}\left\{ i\omega_n f_s^{+
\ast}(t'(0,\alpha))e^{-is\alpha'(0,\alpha)}-\partial_t\left\{
f_s^{+ \ast}(t'(t,\alpha))e^{-is\alpha'(t,\alpha)} \right\}\left|_{t=0}\right.
\right\}
\label{4.4}
\end{equation}
where
\begin{eqnarray}
\alpha'(t,\alpha)=\tan^{-1}\left(\frac{\cosh(\frac{\lambda  t}{2})
\sin(\alpha)}{\cosh(\frac{\lambda
t}{2})\cos(\alpha)\cosh(\tau)-\sinh(\frac{\lambda
t}{2})\sinh(\tau)}
\right)  \nonumber \\
t'(t,\alpha)=\frac{2}{\lambda }\sinh^{-1}\left(\sinh(\frac{\lambda
t}{2})\cosh(\tau)-\cosh(\frac{\lambda
t}{2})\cos(\alpha)\sinh(\tau) \right) \label{4.5}
\end{eqnarray}
and
\begin{equation}
\tau=\frac{\lambda }{2}(T'_0-T_0)
\label{4.6}
\end{equation}

\section{Total number of particles created}
To find out whether the total number of particles created is finite
we must find out if $\beta(n,s)$ is Hilbert-Schmidt, namely
\begin{equation}
\sum_s \sum_n \left| \beta(n,s) \right|^2 < \infty.
\label{5.1}
\end{equation}
If this inequality holds it means that the total number of
created particles is finite and the Bogolubov
transformation is unitarily implementable. To calculate the total
number of particles created we write $\beta(n,s)$ in a slightly different form,
\begin{equation}
\beta(n,s)=\frac{-i}{4\pi\sqrt{\omega_s\omega_n}}\int\! d\alpha
e^{-i(n+s)\alpha}
e^{-is(\alpha'(0,\alpha)-\alpha)}g(n,s,\alpha)
\label{5.2}
\end{equation}
where
\begin{equation}
g(n,s,\alpha)=\left\{ i\omega_n f_s^{+\ast}(t'(0,\alpha))
-is\frac{\partial \alpha'}{\partial t}f_s^{+\ast}(t'(t,\alpha)) -\frac{\partial
t'}{\partial t}\partial_{t'}\left(  f_s^{+\ast}(t'(t,\alpha))\right)
\right\}\left|_{t=0}\right.
\label{5.3}
\end{equation}
We have written $\beta(n,s)$ is this form
allow us to write $\alpha'(0,\alpha)$ in a form which takes care
of the problem of which branch of the $\tan^{-1}(y)$ in (\ref{4.5}) to take.

To investigate the
asymptotic form of $g(n,s,\alpha)$ we have to find the asymptotic behaviour
of the hypergeometric functions involved. The first simplification that
can be made is due to the fact that $\beta=\alpha^{\ast}$. By writing
$\alpha=a+ib$ we see directly from the series for the hypergeometric functions
that for large $a$ one can drop the imaginary part
of $\alpha$
\begin{eqnarray}
H(\alpha,\beta,c,z)&=& 1+\frac{\alpha\beta}{c}z+
\frac{\alpha\beta(\alpha+1)(\beta+1)}{c(c+1)}\frac{z^2}{2}+...
\nonumber \\
&=& 1 + \frac{(a^2+b^2)}{c}z+
\frac{(a^2+b^2)((a+1)^2+b^2)}{c(c+1)}\frac{z^2}{2}+...
\label{5.4} \\
&\approx&1+\frac{(a^2)}{c}z+
\frac{(a^2)(a+1)^2}{c(c+1)}\frac{z^2}{2}+... \nonumber  \\
&=& H(a,a,c,z) \nonumber
\end{eqnarray}
{}From (\ref{2.6}) we see that we need asymptotic forms for hypergeometric
functions of the form $H(a,a,{1\over2},-x^2)$ and $x H(a+{1\over 2},b+{1\over
2},{3\over2},-x^2)$. For the first form we can write the hypergeometric
function
in terms of a Legendre function using the identity \cite{abram}
\begin{equation}
H=(a,a,1/2,-x)={\pi^{{-1\over 2}} \over 2}\Gamma(a+{1\over 2})\Gamma(1-a)
(1+x)^{-a}\left( P_{2a-1}[{x^{1\over 2}\over\sqrt{(1+x)}}]+P_{2a-1}[{-x^{1\over
2}\over\sqrt{(1+x)}}] \right)
\label{5.5}
\end{equation}
To obtain the asymptotic form for $x H(a+{1\over 2},b+{1\over 2},3/2,-x^2)$
we notice that we can write it in terms of the derivative of the first
hypergeometric function,
\begin{equation}
x H(a+{1\over 2},a+{1\over 2},{3\over 2},-x^2)={-1\over 2(a-{1\over
2})^2}\frac{d}{dx}H(a-{1\over 2},a-{1\over 2},1/2,-x^2)
\label{5.6}
\end{equation}
We now use an expression for the Legendre functions valid for large $\nu$
\cite{abram2},
\begin{equation}
P_\nu[\cos(\theta)]\approx{\Gamma(\nu +1)\over \Gamma(\nu +{3\over 2} )}
\sqrt{ 2 \over \pi \sin(\theta)}\cos((\nu+{1\over 2})\theta-{\pi\over 4})
\label{5.7}
\end{equation}
 By using the reflection formula
\begin{equation}
\Gamma(1-x)=\frac{\pi}{\Gamma(x)\sin(\pi z)}
\label{5.8}
\end{equation}
and taking the asymptotic form for the gamma functions which is valid for
large argument,
\begin{equation}
\Gamma(ax+b)\approx \sqrt{2\pi}e^{-ax}(ax)^{ax+b-\frac{1}{2}}
\label{5.9}
\end{equation}
we find that the gamma functions from (\ref{5.5})  and (\ref{5.7}) combine in
such a way as to cancel their $s$ dependence leaving,
\begin{eqnarray}
\beta(n,s)& &=\frac{1}{4\pi\sqrt{\omega_s\omega_n}}\int\!\! d\alpha
e^{-i(n+s)\alpha}e^{-is(\alpha'(0,\alpha)-\alpha)}
\left(\cos[s\cos^{-1}[p(\alpha)]]+\sin[s\cos^{-1}[p(\alpha)]] \right) \nonumber
\\
&\times&\left( A(\alpha)\left(\cos[\frac{\pi s}{2}]+\sin[\frac{\pi s}{2}]
\right)+B(\alpha)\left(\cos[\frac{\pi s}{2}]-\sin[\frac{\pi s}{2}] \right)
\right)
\label{5.10}
\end{eqnarray}
where
\begin{equation}
p(\alpha)=  \frac{\cos(\alpha){\rm sinh}(\tau)}{
\sqrt{1+\cos^2(\alpha){\rm sinh^2}(\tau)}}
\label{5.11}
\end{equation}
\begin{eqnarray}
A(\alpha)&=&\frac{\lambda\left|s\right|}{f(\alpha)^2
(2\left|s\right|-1)^2}\left\{
f(\alpha)^2\left|n\right|(1-(-1)^s)(\left|s\right|-1)+
i\cos[\alpha]\sinh[\tau](1-(-1)^s) \right. \nonumber \\
&+& \left.
i\cosh[\tau](1+(-1)^s)(2-f(\alpha)+f(\alpha)\left|s\right|-f(\alpha)s^2)
\right. \nonumber \\
&+& \left.
\sin[\alpha]\sinh[\tau](1-(-1)^s)(f(\alpha)s\left|s\right|-f(\alpha)s)
\right\}
\label{5.12}
\end{eqnarray}
\begin{equation}
B(\alpha)=\frac{\lambda}{4f(\alpha)}\left\{if(\alpha)
\left|n\right|(1+(-1)^s))+\left|s\right|\cosh[\tau](1-(-1)^s)
+is\sin[\alpha]\sinh[\tau](1+(-1)^s)   \right\}
\label{5.13}
\end{equation}
and
\begin{equation}
f(\alpha)= 1 + \cos^2[\alpha]\sinh^2[\tau]
\label{5.14}
\end{equation}
The entire point of writing $\beta(n,s)$ in this way was to allow
us to integrate the above expression by parts. After expanding the
$\sin[s\cos^{-1}[p(\alpha)]]$ and $\cos[s\cos^{-1}[p(\alpha)]]$ in
terms of exponentials, each term making up $\beta(n,s)$ can be written
in the form,
\begin{equation}
\beta(n,s)\propto \int d\alpha e^{-i(n+s)\alpha}e^{isg(\alpha)}K_{n,s}(\alpha)
\label{5.15}
\end{equation}
where $K_{n,s}(\alpha)$ incorporates the last term in (\ref{5.10}) which
contains $A(\alpha)$ and $B(\alpha)$ and
\begin{equation}
g(\alpha)=-(\alpha'(0,\alpha)-\alpha)\pm \cos^{-1}(p(\alpha))
\label{5.16}
\end{equation}
where the $\pm$ depends on which of the two terms one is dealing with.
The important point is that the behaviour of $K_{n,s}(\alpha)$ in terms of
$n,s$
does not change because the dependence on $n,s$ is decoupled from $\alpha$.
not increase if one differentiates it with respect to $\alpha$. One can now
integrate by parts indefinitely to observe that the expression must drop
off faster than any inverse power of $n,s$. For example after integrating
by parts twice one is left with,
\begin{equation}
\beta(n,s)\propto \int d\alpha  e^{-i(n+s)\alpha} e^{\pm
isg(\alpha)}\frac{d}{d\alpha}\left(
\frac{1}{-i(n+s)\pm
g'(\alpha)}\frac{d}{d\alpha}\left(\frac{K_{n,s}(\alpha)}{-i(n+s)\pm g'(\alpha)}
   \right) \right)
\label{5.17}
\end{equation}
The only problem that could arise is if $-(n+s)-g'(\alpha)$ ever vanished.
This is not a problem however because we are interested in the large momenta
limit and $g'(\alpha)$ is a well behaved function.

\section{Conclusions}
We have shown that $\beta(n,s)$ drops off faster than any inverse
power of $n,s$, for large $n,s$. This implies that the total number of
particles
created is finite and therefore the Bogolubov transformation is unitarily
implementable. The fact that the total number of particles
created is finite is in agreement with Fulling's analysis for an isotropic
universe of finite spatial volume \cite{fulling}.

If in fact $\beta(n,s)$ drops off like an exponential then after performing one
of the sums in $\left|\beta(n,s) \right|^2$ one will be left with a Planck
spectrum. This is to be expected as for large momenta our analysis should be
similar
in nature to the analysis of massless particle creation.

It should be emphasized that this calculation does not involve calculating the
Bogolubov transformation relating in essence to different spacetimes. This
calculation involves comparing an observer's particle definition at two
different
times in the same spacetime. In this way one is not misinterpreting boundary
effects by comparing fields quantized in overlapping but different spacetimes
\cite{1.2} as is the case in the standard Rindler analysis and many other
calculations.

\section{Acknowledgements} This research was
supported in part by a grant from the Natural Sciences and
Engineering Research Council of Canada.

\end{document}